# Meixner - Pollaczek Quantum System

T. J. Taiwo

*Department of Mathematics, University of Lagos. Akoka. Lagos- State. P. O. Box 101017, Nigeria*

**Abstract**: In an effort to achieve our aims (getting larger quantum systems) in the recent reformulation of quantum mechanics without potential function [1-5], we obtained a new quantum system associated with Meixner - Pollaczek orthogonal polynomial class (**hereby called Meixner - Pollaczek Quantum system**). The energy spectrum and wavefunction of the quantum system were given. To the best of our knowledge, this quantum system is not found in any physics literature.

## 1 Introduction

The connection between scattering and the asymptotics of orthogonal polynomials were described by Case and Geronimo [6-8]. As a result of these, a recent reformulation of quantum mechanics without potential function [1-5] was formulated. Basically, the objective was to obtain a set of analytically realizable systems, which is larger than in the standard formulation and that may or may not be associated with any given or previously known potential functions. In this reformulation we have shown that the wavefunction associated with any physical system can be written as

$$\psi_E^\mu(x) = \sum_n f_n^\mu(E)\phi_n(x) \qquad (1)$$

where $\{\mu\}$ is a set of parameters associated with the physical system, $\{f_n^\mu(E)\}$ is parameterized function of the energy, and $\{\phi_n^\mu(x)\}$ is set of basis set. Writing $f_n^\mu(E) = f_0^\mu(E) P_n^\mu(\varepsilon)$; by the completeness of the basis and energy normalization of the density of state make $\{P_n^\mu(\varepsilon)\}$ a complete set of orthogonal polynomials with positive weight function $\rho^\mu(\varepsilon) = \left[f_0^\mu(E)\right]^2$. Hence we have

$$\int \rho^\mu(\varepsilon) P_n^\mu(\varepsilon) P_m^\mu(\varepsilon) d\zeta(\varepsilon) = \delta_{nm} \qquad (2)$$

where $\varepsilon$ is proper function of $E$ and $\{\mu\}$; and $d\zeta(\varepsilon)$ is an appropriate energy integration measure. Therefore (1) becomes

$$\psi_E^\mu(x) = \sqrt{\rho^\mu(\varepsilon)} \sum_n P_n^\mu(\varepsilon)\phi_n(x) \qquad (3)$$

Also, we expect all relevant orthogonal polynomials to have asymptotics $(n \to \infty)$ behaviour

$$P_n^\mu(\varepsilon) \approx n^{-\tau} A^\mu(\varepsilon) \cos\left[n^\xi \theta(\varepsilon) + \delta^\mu(\varepsilon)\right] \qquad (4)$$

where $\tau$ and $\xi$ are real positive constants. $A^\mu(\varepsilon)$ is the scattering amplitude and $\delta^\mu(\varepsilon)$ is the phase shift. Bound states occur as a set (finite or infinite) of zeros of the energies that make the scattering amplitude vanish. So, for a $k^{\text{th}}$ bound state at an energy $E_k = E(\varepsilon_k)$ such that $A^\mu(\varepsilon_k) = 0$; the bound state wavefunction becomes

$$\psi_k^\mu(x) = \sqrt{\omega^\mu(\varepsilon_k)} \sum_n Q_n^\mu(\varepsilon_k)\phi_n(x) \qquad (5)$$

$\{Q_n^\mu(\varepsilon_k)\}$ are discrete version of the polynomials $\{P_n^\mu(\varepsilon)\}$ and $\omega^\mu(\varepsilon_k)$ is the associated discrete weight function that satisfy $\sum_k \omega^\mu(\varepsilon_k) Q_n^\mu(\varepsilon_k) Q_m^\mu(\varepsilon_k) = \delta_{n,m}$. Also we can have the wavefunction written as

$$\psi_k^\mu(x,E) = \sqrt{\rho^\mu(\varepsilon)} \sum_n P_n^\mu(\varepsilon) \phi_n(x) + \sqrt{\omega^\mu(\varepsilon_k)} \sum_n P_n^\mu(\varepsilon_k) \phi_n(x) \qquad (6)$$

this corresponds to the case where there coexist continuous as well as discrete energy spectra simultaneously. In this reformulation, the set of the orthogonal polynomials plays the role of potential function in conventional quantum mechanics and also carry the kinematic information- angular momentum. So in the absence of a potential function, the physical properties of the system are deduced from the features (weight function, nature of generating function, distribution and density of the polynomial zeros, recursion relation, asymptotics, differential or difference equation) of the orthogonal polynomials $\{P_n^\mu(\varepsilon), Q_n^\mu(\varepsilon_m)\}$.

Also the orthogonal polynomials (continuous and discrete) satisfy three term recursion relations, and produce a tridiagonal matrix representation for the wave operator through the basis set $\{\phi_n^\mu(x)\}$. As a result, the matrix wave equation $H|\psi\rangle = E|\psi\rangle$ become equivalent to $H|P_n^\mu(\varepsilon)\rangle = E\Omega|P_n^\mu(\varepsilon)\rangle$, where $\Omega_{nm} = \langle \phi_n | \phi_m \rangle$, which should yield the recursion relation $\varepsilon P_n^\mu(\varepsilon) = a_n^\mu P_n^\mu(\varepsilon) + b_{n-1}^\mu P_{n-1}^\mu(\varepsilon) + b_n^\mu P_{n+1}^\mu(\varepsilon)$, for $n=1,2,..$, where $P_0^\mu(\varepsilon) = 1$, $P_1^\mu(\varepsilon) = \alpha\varepsilon + \beta$ and $\{a_n^\mu, b_n^\mu\}$ are the recursion coefficients with $b_n^\mu \neq 0$ for all $n$.

In [2], physical systems that belong to two parameters (three parameters) Meixner - Pollaczek polynomial (continuous dual Hahn polynomial) have been studied. Conventional quantum systems, such as, Coulomb, oscillator, and Morse, were obtained. In addition, new systems that do not belong to known class of exactly solvable problems were found. Furthermore, in [3], we obtained another quantum systems associated with the four parameters Wilson- Racah Orthogonal polynomials were deduced also.

Since we expect these systems to have potential function, procedures to get them were formulated in [5]. Basically, four formulas were introduced with different degree of accuracy. We have used these formulas and obtained reliable results (potential functions) associated with known exactly solvable quantum system and the unknown ones gotten in this formulation

Here, we consider a new quantum system associated with Meixner- Pollaczek Orthogonal polynomial class. This quantum system has not been treated in physics literature to the best of our knowledge. We obtain the energy spectrum and wavefunction of this system. We named this quantum system as the **Meixner Pollaczek Quantum system.** This quantum system has intrinsic property, which is, the potential function is energy dependent and will be of greater application in condensed matter physics or material sciences.

## 2. The Meixner - Pollaczek Quantum System

This new quantum system is defined on the real line with basis element $\phi_n(x) = \sqrt{\dfrac{\Gamma(n+1)}{\Gamma(n+\nu+1)}} (\lambda x)^{\frac{\nu}{2}} e^{\lambda x/2} L_n^\nu(\lambda x)$, where $L_n^\nu(y)$ is a Laguerre polynomial in the configuration space $y = \lambda x$ for $x \in [0,\infty]$ and $\nu > -1$. The associated orthogonal polynomial is the **two –parameters Meixner - Pollaczek** orthogonal polynomial defined as

$$P_n^\mu(z,\theta) = \sqrt{\dfrac{(2\mu)_n}{n!}} e^{in\theta} {}_2F_1\left(\begin{matrix}-n,\mu+iz\\ 2\mu\end{matrix}\bigg| 1-e^{-2i\theta}\right) \qquad (7)$$

where $(a)_n = a(a+1)(a+2)...(a+n-1) = \dfrac{\Gamma(n+a)}{\Gamma(a)}$, $z$ is the whole real line, $\mu > 0$ and $0 < \theta < \pi$. The orthogonality relation for this polynomial is

$$\int_{-\infty}^{\infty} \rho^{\mu}(z,\theta) P_n^{\mu}(z,\theta) P_m^{\mu}(z,\theta) dz = \delta_{nm} \tag{8}$$

where $\rho^{\mu}(z,\theta) = \frac{1}{2\pi\Gamma(2\mu)} (2\sin\theta)^{2\mu} e^{(2\theta-\pi)z} |\Gamma(\mu+iz)|^2$ is the normalized weight function. The three term recursive relation for this polynomial

$$(z\sin\theta) P_n^{\mu}(z,\theta) = -\left[(n+\mu)\cos\theta\right] P_n^{\mu}(z,\theta)$$
$$+ \frac{1}{2}\sqrt{n(n+2\mu-1)} P_{n-1}^{\mu}(z,\theta) + \frac{1}{2}\sqrt{(n+1)(n+2\mu)} P_{n+1}^{\mu}(z,\theta) \tag{9}$$

In [2- Appendix A], it was shown that the asymptotics of this polynomial is

$$P_n^{\mu}(z,\theta) \approx \frac{2n^{-1/2} e^{\left(\frac{\pi}{2}-\theta\right)z}}{(2\sin\theta)^{\mu} \Gamma(\mu+iz)} \cos\left[n\theta + \arg\Gamma(\mu+iz) - \mu\frac{\pi}{2} - z\ln(2n\sin\theta)\right] \tag{10}$$

Since $\ln n \approx o(n)$ as $n \to \infty$, we can neglect $z\ln(2n\sin\theta)$ relative to $n\theta$. Comparing (10) to (4), the scattering amplitude is

$$A^{\mu}(\varepsilon) = \frac{2e^{\left(\frac{\pi}{2}-\theta\right)z}}{(2\sin\theta)^{\mu} \Gamma(\mu+iz)} \tag{11}$$

and phase shift is

$$\delta^{\mu}(\varepsilon) = \arg\Gamma(\mu+iz) \tag{12}$$

From the scattering amplitude (11), discrete infinite spectrum occurs if $\mu+iz = -m$, where $m = 0,1,2,...,N$. Thus the spectrum formula associated with this polynomial is $z_m^2 = -(m+\mu)^2$. For the **Meixner- Pollaczek quantum system**, we choose our physical parameters as:

$$z = \ln(\kappa/\lambda), \cos\theta = \frac{\kappa-\mu\lambda}{\kappa+\mu\lambda}, \mu = \rho+1, \nu = \ell+1, \text{ and } \kappa^2 = 2E; \tag{13}$$

where $\rho > -1$ and $\ell > -1$

Therefore, the energy spectrum for the bound states is

$$E_m = \frac{\lambda^2}{2} e^{2(m+\mu)} = \frac{\lambda^2}{2} e^{2(m+\rho+1)} \tag{14}$$

and scattering phase shift

$$\delta^{\mu}(\varepsilon) = \arg\Gamma(\rho+1+i\ln(\kappa/\lambda)) \tag{15}$$

The discrete infinite bound state wavefunction can be obtained using (5) and where the expansion polynomial coefficient will be the **discrete version of Meixner Pollaczek polynomial** obtained by substituting $z = i(m+\mu)$ and $\theta \to i\theta$ in (7) to get

$$M_n^{\mu}(m;\beta) = \sqrt{\frac{(2\mu)_n}{n!}} \beta^{n/2} {}_2F_1\left(\begin{array}{c}-n,-m\\2\mu\end{array}\bigg|1-\beta^{-1}\right) \tag{16}$$

where $\beta = e^{-2\theta}$ with $\theta > 0$ making $0 < \beta < 1$. With this substitution and denoting $2\cosh\theta = \frac{1}{\sqrt{\beta}} + \sqrt{\beta}$ and $2\sinh\theta = \frac{1}{\sqrt{\beta}} - \sqrt{\beta}$, then the recursive relation (9) becomes

$$(\beta-1)mM_n^{\mu}(m;\beta) = -\left[n(1+\beta) + 2\mu\beta\right] M_n^{\mu}(m;\beta)$$
$$+ \sqrt{n(n+2\mu-1)\beta} M_{n-1}^{\mu}(m;\beta) + \sqrt{(n+1)(n+2\mu)\beta} M_{n+1}^{\mu}(m;\beta) \tag{17}$$

for the discrete Meixner Pollaczek polynomial with normalized discrete weight function $\omega_m^\mu(\beta) = (1-\beta)^{2\mu} \frac{\Gamma(m+2\mu)}{\Gamma(2\mu)\Gamma(m+1)}$, and orthogonality $\sum_{m=0}^{\infty} \omega_m^\mu(\beta) M_n^\mu(m;\beta) M_\alpha^\mu(m;\beta) = \delta_{n\alpha}$. Now suppose we take $2\mu = -N$, where $N$ is a non-negative integer and $\beta = \frac{-\gamma}{1-\gamma}$, then (16) will become the normalized discrete **Krawtchouk polynomial (finite spectrum)** defined as

$$K_n^\mu(m;\gamma) = \sqrt{\frac{N!}{n!(N-n)!}} \left(\frac{\gamma}{1-\gamma}\right)^{n/2} {}_2F_1\left(\begin{matrix}-n,-m\\-N\end{matrix}\bigg|\gamma^{-1}\right) \tag{18}$$

where $\gamma^{-1} = 1 - \beta^{-1}$ with $0 < \gamma < 1$ and $n, m = 0, 1, ..., N$. Note that we have used $(-N)_n = \frac{\Gamma(n-N)}{\Gamma(-N)} = (-1)^n \frac{\Gamma(N+1)}{\Gamma(N-n+1)}$ in getting (18) from (16). The recursive relation this polynomial will be

$$mK_n^\mu(m;\beta) = \left[N\gamma + n(1-2\gamma)\right] K_n^\mu(m;\gamma) \tag{19}$$
$$-\sqrt{n(N-n+1)\gamma(1-\gamma)} K_{n-1}^\mu(m;\gamma) + \sqrt{(n+1)(N-n)\gamma(1-\gamma)} K_{n+1}^\mu(m;\gamma)$$

with discrete normalized weight function $\omega_m^N(\gamma) = (1-\gamma)^{N-m} \frac{\Gamma(N+1)\gamma^m}{\Gamma(N-m+1)\Gamma(m+1)}$, and orthogonality $\sum_{m=0}^{N} \omega_m^N(\gamma) K_n^N(m;\gamma) K_\alpha^N(m;\gamma) = \delta_{n\alpha}$.

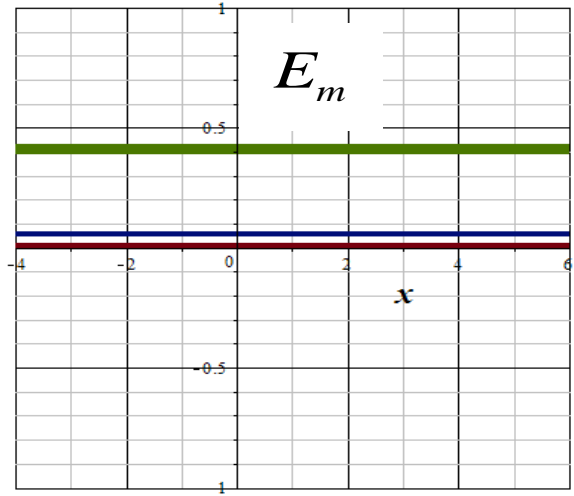

**FIG.1**. The bound states energy spectrum (14) for the new quantum system with physical parameters: $\lambda = 0.1$, $\mu = 0.2$, and $m = 0, 1, 2, ...$,

## 3. Energy Spectrum and Wavefunction of the Meixner - Pollaczek Quantum System

To further confirm our work, we sought to get the energy spectrum numerically and thereafter plot the wave function. Now, substituting the physical parameters (13) in (9), we have

$$\left[2b(n+\rho+1) - 4\sqrt{\kappa b}\left(\ln\frac{\kappa}{\lambda}\right)\right] P_n^\mu(z,\theta) + b\left(\sqrt{n(n+2\rho+1)} P_{n-1}^\mu(z,\theta) + \sqrt{(n+1)(n+2\rho+2)} P_{n+1}^\mu(z,\theta)\right)$$
$$= \kappa\left(2(n+\rho+1) P_n^\mu(z,\theta) - \sqrt{n(n+2\rho+1)} P_{n-1}^\mu(z,\theta) - \sqrt{(n+1)(n+2\rho+2)} P_{n+1}^\mu(z,\theta)\right) \tag{20}$$

where $b = \mu\lambda$. We can rewrite (20) as $H|P\rangle = \varepsilon|P\rangle$ where $\varepsilon = \kappa = \sqrt{2E}$. If we compare (20) to $H|P_n^\mu(\varepsilon)\rangle = E\Omega|P_n^\varepsilon(\varepsilon)\rangle$, we obtained the Hamiltonian matrix element as

$$H_{n,m} = \left[2b(n+\rho+1) - 4\sqrt{\kappa b}\left(\ln\frac{\kappa}{\lambda}\right)\right]\delta_{nm} + b\left(\sqrt{n(n+2\rho+1)}\delta_{n,m+1} + \sqrt{(n+1)(n+2\rho+2)}\delta_{n,m-1}\right) \quad (21)$$

where $\Omega_{nm} = 2(n+\rho+1)\delta_{nm} - \sqrt{n(n+2\rho+1)}\delta_{n,m+1} - \sqrt{(n+1)(n+2\rho+2)}\delta_{n,m-1}$. We observed that the Hamiltonian matrix element is energy dependent. So in getting the potential function of the **Meixner -Pollaczek** quantum system will have to be done using a numerically approach. However, this will be for another time. Similarly, we get the three –term recursion relation for the expansion coefficient of the wave function as

$$z'P_n^\mu(z,\theta) = 2b(n+\rho+1)P_n^\mu(z,\theta) + b\sqrt{n(n+2\rho+1)}P_{n-1}^\mu(z,\theta) + b\sqrt{(n+1)(n+2\rho+2)}P_{n+1}^\mu(z,\theta) \quad (22)$$

where $z' = 4\sqrt{\kappa b}\left(\ln\frac{\kappa}{\lambda}\right)$. Writing $f_n(E) = f_0(E)P_0(E)$, we make $P_0(E) = 1$. Now when $n = 0$; will get

$$P_1 = \frac{(z' - a_0)}{c_0},$$ where $a_0 = 2b(\rho+1)$, and $c_0 = b\sqrt{(2\rho+2)}$. This relation is valid for $n = 0,1,2,...$,

Now we obtain the Hamiltonian matrix from (21) as $H = J|_{E=0}$. Then, the energy spectrum is calculated from the wave equation $H|\psi\rangle = E|\psi\rangle$ as the generalized eigenvalues $\{E\}$ of the matrix equation $\sum_m H_{n,m} f_m = E\sum_m \Omega_{n,m} f_m$; where $\Omega_{nm} = 2(n+\rho+1)\delta_{nm} - \sqrt{n(n+2\rho+1)}\delta_{n,m+1} - \sqrt{(n+1)(n+2\rho+2)}\delta_{n,m-1}$ is the overlap basis element.

Table 1 is a list of the lowest energy spectrum for a given set of values of the physical parameters and for various basis sizes. We show only significant decimal digits that do not change with any substantial increase in the basis size. One intrinsic feature we observed was that there are some ranges within the eigenvalues that stable result was gotten – the wave function obtained was same despite the variation in the eigenvalues. Also, across the negative (orange color) values of the eigenvalue, the wavefunction was not gotten due to the energy parameter function $z'$. As we progress to the positive value (blue color along the table); wave function is defined. And as we move down the table (tan color); we obtained a shift in the wavefunction. This phenomenon was same using across each column of the table even as the basis size increases.

In Figure 2, we plot the bound state wavefunctions corresponding to the physical configuration and energy spectrum of Table 1. We calculate the $m^{\text{th}}$ bound state using the sum $\psi(E_m, x) \sim \sum_{n=0}^{N-1} P_n(\varepsilon_m)\phi_n(x)$, where $N$ is some large enough integer. We could have used (6) in plotting the wave function had it been a well - know polynomial.

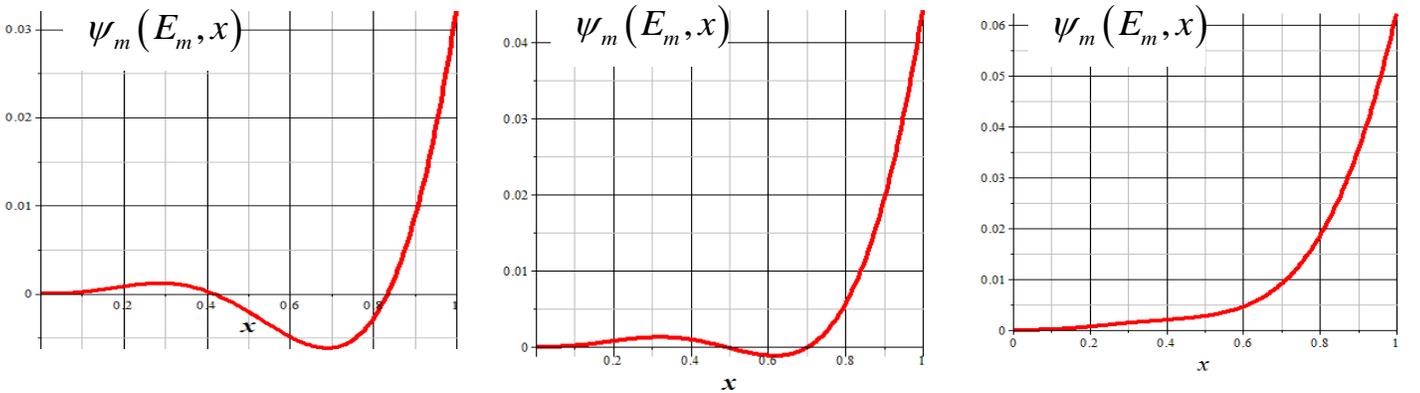

**FIG.2**. The graph of the wavefunction as we move across the table from the positive eigenvalues for the new quantum system with physical parameters: $\lambda = 0.7$, $\mu = 1.1$, $\rho = 2.1$ $b = 0.77$, and $v = 0.3$.

Table 1: The eigenvalues of the generalized wave equation for physical parameter: $\lambda = 0.7$, $\mu = 1.1$, $\rho = 2.1$ $b = 0.77$, and $v = 0.3$.

| n | 10 by 10 | 20 by 20 | 50 by 50 |
|---|---|---|---|
| 0 | -0.3182625110 | -0.3539806785 | -0.3737970285 |
|   | -0.2704316853 | -0.3377886422 | -0.3697824370 |
|   | -0.2034360814 | -0.3181879752 | -0.3654326406 |
|   | -0.1066454946 | -0.2940367086 | -0.3605803981 |
|   | 0.0377168281 | -0.2641948959 | -0.3551371874 |
|   | 0.2632400773 | -0.2272618962 | -0.3490336853 |
|   | 0.6407379301 | -0.1813997506 | -0.3422055319 |
|   | 1.3431303711 | -0.1241121585 | -0.3345878470 |
|   | 2.9008353557 | -0.0519168187 | -0.3261122685 |
|   | 7.8219861758 | 0.0401731042 | -0.3167048558 |
|   |   | 0.1595076545 | -0.3062842782 |
|   |   | 0.3172869557 | -0.2947600362 |
|   |   | 0.5312685097 | -0.2820305748 |
|   |   | 0.8309994740 | -0.2679811864 |
|   |   | 1.2686957529 | -0.2524816114 |
|   |   | 1.9440545704 | -0.2353832386 |
|   |   | 3.0681761380 | -0.2165157939 |
|   |   | 5.1586932727 | -0.1956833791 |
|   |   | 9.8117373105 | -0.1726596925 |
|   |   | 24.6289177244 | -0.1471822153 |
|   |   |   | -0.1189450864 |
|   |   |   | -0.0875903080 |
|   |   |   | -0.0526968083 |
|   |   |   | -0.0137667391 |
|   |   |   | 0.0297918277 |
|   |   |   | 0.0786869331 |
|   |   |   | 0.1337710351 |
|   |   |   | 0.1960777130 |
|   |   |   | 0.2668697536 |
|   |   |   | 0.3477028418 |
|   |   |   | 0.4405109317 |
|   |   |   | 0.5477221710 |
|   |   |   | 0.6724185590 |
|   |   |   | 0.8185592905 |
|   |   |   | 0.9912985971 |
|   |   |   | 1.1974467752 |
|   |   |   | 1.4461532792 |
|   |   |   | 1.7499433240 |
|   |   |   | 2.1263340792 |
|   |   |   | 2.6004336118 |
|   |   |   | 3.2092718167 |
|   |   |   | 4.0093243214 |
|   |   |   | 5.0902438596 |
|   |   |   | 6.6014543295 |
|   |   |   | 8.8075665036 |
|   |   |   | 12.2150693720 |
|   |   |   | 17.8995854711 |
|   |   |   | 28.5079491893 |
|   |   |   | 52.2371825915 |
|   |   |   | 128.3492106634 |

## 4. Conclusion

In this paper, we derived analytically the **Meixner - Pollaczek quantum system**. The energy spectrum, scattering phase shift, and wavefunction of the system were shown. However, it was observed the potential function of this system is energy dependent which can only be found using a numerically approach. We will therefore follow this work later in deriving the potential function.

## Acknowledgement

The author highly appreciates the support of the Saudi Centre for Theoretical Physics during the progress of this work. Specifically, this paper is dedicated to **Prof. A.D. Alhaidari** for his fatherly role and support in impacting knowledge. Allah in His infinite mercy will continue to bless him more.

## References:

[1] A.D. Alhaidari, *Formulation of quantum mechanics without potential function*. Quant. Phys. Lett. **4** (2015) 51

[2] A.D. Alhaidari and M.E.H. Ismail, *Quantum mechanics without Potential Function*, J. Math. Phys. **56** (2015) 072107

[3] A.D. Alhaidari and T. J. Taiwo, *Wilson –Racah quantum system*, J. Math. Phys. **58** (2017) 022101

[4] T.J. Taiwo, "New Quantum System of the Wilson Orthogonal Polynomial", https://arxiv.org/abs/1709.08896. Submitted

[5] A.D. Alhaidari, *Establishing correspondence between the reformulation of quantum mechanics without a potential function and the conventional formulation*. Arxiv.org/abs/1703.08659. Submitted.

[6] K. M. Case, "Orthogonal polynomials from the view point of scattering theory," J. Math. Phys. **15,** 2166 -2174 (1974)

[7] J.S. Geronimo and K. M. Case, " Scattering theory and polynomials orthogonal on real line," Trans. Am . Math. Soc **258**, 467 – 494 (1980)

[8] J. S. Geronimo, " A relation between the coefficients in the recurrence formula and spectral function for Orthogonal polynomials," Trans. Am. Math. Soc. **260,** 65 – 82 (1980)

[9] R.W. Haymaker and L. Schlessinger, *The Pade Approximation in Theoretical Physics*, edited by G.A. Baker and J.L Gammel (Academic Press, New York, 1970)

[10] R. Koekoek and R. Swarttouw, *The Askey –Scheme of Hypergeometric orthogonal polynomials and its q analogues,* Reports of the Faculty of Technical Mathematics and Informatics, Number 98 -17 (Delft University of Technology, Delft, 1998)

[11] A.D. Alhaidari, *Solution of the nonrelativistic wave equation using the Tridiagonal representation approach,* J. Math. Phys. **58** (2017) 072104.

[12] A.D. Alhaidari and H. Bahlouli, *Extending the class of solvable potentials I. The infinite potential well with a sinusoidal bottom,* J. Math. Phys. 49 (2008) 082102

[13] J.A. Wilson, *Asymptotic for the $_4F_3$ polynomials*, J. Approximation Theory **66,** 58 -71 (1991)

[14] F.W.J. Oliver, *Asymptotic and Special Functions*, (Academic Press, New York, 1974)